\documentclass[pra,twocolumn]{revtex4-1}

\usepackage{amsmath}
\usepackage{graphicx}

\begin{document}

\title{Continuous-variable dense coding via a general Gaussian state: Monogamy relation}

\author{Jaehak Lee}
\author{Se-Wan Ji}
\author{Jiyong Park}
\author{Hyunchul Nha}
\affiliation{Department of Physics, Texas A \& M University at Qatar, P.O. Box 23874, Doha, Qatar}

\begin{abstract}
We study a continuous variable (CV) dense-coding protocol, originally proposed to employ a two-mode squeezed state, using a general two-mode Gaussian state as a quantum channel. We particularly obtain conditions to manifest quantum advantage by beating two well-known single-mode schemes, namely, the squeezed-state scheme (best Gaussian scheme) and the number-state scheme (optimal scheme achieving the Holevo bound). We then extend our study to a multipartite Gaussian state and investigate the monogamy of operational entanglement measured by the communication capacity under the dense-coding protocol. 
We show that this operational entanglement represents a strict monogamy relation, by means of Heisenberg's uncertainty principle among different parties, i.e., the quantum advantage for communication can be possible for only one pair of two-mode systems among many parties.
\end{abstract}

\maketitle

\section{\label{sec:introduction}Introduction}

% entanglement for quantum information protocol
Quantum correlation, especially entanglement \cite{RevModPhys.81.865}, is a key resource for quantum information processing, e.g., quantum teleportation \cite{bib:PhysRevLett.70.1895, bib:PhysRevLett.80.869} and quantum dense coding \cite{bib:PhysRevLett.69.2881, bib:PhysRevA.61.042302}. Performance in such protocols, quantified by the output fidelity of teleportation and the communication capacity of dense coding, respectively, can be used as an operational measure of entanglement. When entanglement is shared by more than two parties, we may look into the multipartite entanglement structure and its usefulness by selecting a few parties among all and investigating their performance in quantum protocols for all such selected subsystems.

% monogamy relation for entanglement
Numerous studies have so far demonstrated that a monogamy relation is one of the fundamental properties of multipartite entanglement, that is, entanglement between some subsystems puts limitation on their correlation with the other parties. For instance, if two quantum systems are maximally entangled, they cannot be correlated with a third party at all, even classically \cite{bib:PhysRevA.61.052306}. 
The monogamy relation has been formulated quantitatively with a proper measure of entanglement for discrete variables \cite{bib:PhysRevA.61.052306,bib:PhysRevLett.96.220503}, which was also extended to Gaussian CV systems \cite{bib:NewJPhys.8.15,bib:PhysRevA.73.032345}.
% monogamy relation in quantum communication protocol
On the other hand, only a few studies have attempted to find such monogamy relations in view of {\it useful} entanglement for quantum communication protocols. In Ref. \cite{bib:PhysRevA.79.054309}, it was shown that teleportation fidelity satisfies a monogamy relation for three-qutrit pure states and for $N$-qubit pure states, but not in general. Recently, it was proved that a strict monogamy relation exists in dense coding for discrete variables \cite{bib:PhysRevA.87.052319}, i.e., if a sender has some quantum advantage in dense coding with one receiver, (s)he can not have quantum advantage with any other receivers.

% continuous variable dense coding
In this paper, we study a CV dense coding protocol and show a strict monogamy relation in this regime. Ever since quantum dense coding was first developed in qubit systems \cite{bib:PhysRevLett.69.2881}, the dense coding capacity has been derived for arbitrary finite-dimensional systems \cite{bib:QantumInfComput.1.70} with arbitrary number of parties \cite{bib:PhysRevLett.93.210501} for discrete-variable systems. On the other hand, there have been few studies on CV dense coding with a general Gaussian state. The CV dense coding protocol was first proposed by Braunstein and Kimble \cite{bib:PhysRevA.61.042302} in which a pure two-mode squeezed vacuum (TMSV) state is employed as a quantum resource. A controlled dense coding was later developed \cite{bib:PhysRevA.66.032318}, similar to the network quantum teleportation \cite{bib:PhysRevLett.84.3482}, in which a three-mode entangled state is prepared and information is transferred between two parties with the help of classical communication provided by a third party.

% criterion for quantum advantage in CV dense coding
We here study the dense coding protocol employing an arbitrary two-mode Gaussian state. We then compare this two-mode scheme with two single-mode schemes, namely, the squeezed-state scheme known to be best among Gaussian schemes and the number-state scheme known to be the ultimate optimal scheme achieving the Holevo bound. We derive conditions to beat those two schemes, i.e. to detect two-mode Gaussian states that can manifest quantum advantage in CV dense coding. These conditions can be expressed in terms of the variances of two correlated quadratures as
\begin{equation} \label{eq:criterion}
V_{x_-}V_{p_+} < B_c,
\end{equation}
where $V_{x_-}$ and $V_{p_+}$ are the variances of $x_1-x_2$ and $p_1+p_2$, respectively. 
The bound $B_c$ turns out to be $\left(1/4\right)^2$ and $\left(1/2e\right)^2$ to beat the squeezed-state scheme and the number-state scheme, respectively. 
These are more stringent than the sufficient condition to verify entanglement \cite{bib:PhysRevA.60.2752}, where the bound is $1/4$.

% monogamy relation in CV dense coding
We also extend our study to the case of multipartite Gaussian states. 
We particularly show that, by means of Heisenberg's uncertainty principle, the criterion (\ref{eq:criterion}) can be satisfied for only one pair of two-modes among all. Accordingly, we have a strict monogamy relation for CV dense coding, that is, a sender can have quantum advantage with one receiver only.

\section{\label{sec:description}Gaussian state description}

% covariance matrix
We start with a brief description of Gaussian states and their transformations under Gaussian operations. 

{\it N-mode Gaussian States}---To describe an $N$-mode CV system, we introduce a $2N$-dimensional vector whose components are canonical variables of each mode, given by
\begin{equation}
\xi=(x_1,p_1,x_2,p_2,\cdots,x_N,p_N)^T,
\end{equation}
with their corresponding operators
\begin{equation}
\hat{\xi}=(\hat{x}_1,\hat{p}_1,\hat{x}_2,\hat{p}_2,\cdots,\hat{x}_N,\hat{p}_N)^T.
\end{equation}
A Gaussian state is fully characterized by the first-order moments and the second-order moments of the position and the momentum operators. Its phase-space distribution (Wigner function) takes a form of Gaussian function
\begin{equation}
W = \frac{1}{(2\pi)^N \sqrt{\textrm{det}\boldsymbol{\sigma}}} ~ \textrm{exp} \left[ -\frac{1}{2} (\xi-\bar{\xi})^T \boldsymbol{\sigma}^{-1} (\xi-\bar{\xi}) \right],
\end{equation}
where $\bar{\xi_i}=\langle\hat{\xi}_i\rangle$ are the first-order moments and $\boldsymbol{\sigma}$ is the covariance matrix (CM) whose elements are the second-order moments, 
\begin{equation}
\sigma_{ij} = \frac{1}{2} \langle \hat{\xi}_i\hat{\xi}_j+\hat{\bf \xi}_j\hat{\bf \xi}_i \rangle - \bar{\xi_i}\bar{\xi_j},
\end{equation}
$(i,j=1,\cdots,2N)$. Due to the uncertainty relation arising from $ [\hat{x}_j,\hat{p}_k] = i\delta_{jk}$, every CM must satisfy
\begin{equation} \label{eq:commutation}
\boldsymbol{\sigma} + \frac{i}{2}\Omega_N \geq 0,
~~ \textrm{where } \Omega_N = \bigoplus^N \left( \begin{array}{cc} 0 & 1 \\ -1 & 0 \end{array} \right).
\end{equation}

% symplectic transform
{\it Gaussian Operations}---An arbitrary Gaussian unitary operation is a linear transformation of canonical operators that can be represented as $ \hat{\xi} \to S\hat{\xi} + \boldsymbol{
\lambda} $. Here $\boldsymbol{\lambda}$ is a $2N$-dimensional real vector displacing the first-order moments by $ \bar{\xi} \to \bar{\xi}+\boldsymbol{\lambda} $ and $S$ is a symplectic transformation satisfying $S \Omega_N S^T = \Omega_N $ \cite{bib:JPhys.45.471}. Under a symplectic transformation, CM evolves as $ \boldsymbol{\sigma} \to S \boldsymbol{\sigma} S^T $.

% standard form (arbitrary two mode, pure three mode)
{\it Standard form}---Every two-mode Gaussian state, pure and mixed, can be transformed into a standard form via a set of local symplectic transformations, which can be written as
\begin{equation} \label{eq:cmstandard2}
\boldsymbol{\sigma}_{AB}^\textrm{(s)} =
\left( \begin{array}{cccc}
a_1 & 0 & c_{12} & 0 \\ 0 & a_1 & 0 & d_{12} \\ c_{12} & 0 & a_2 & 0 \\ 0 & d_{12} & 0 & a_2
\end{array} \right) .
\end{equation}
For the case of three-mode systems, a pure three-mode Gaussian state can be transformed to a standard form \cite{bib:PhysRevA.73.032345}
\begin{equation} \label{eq:cmstandard3}
\boldsymbol{\sigma}_{ABC}^\textrm{(s)} =
\left( \begin{array}{cccccc}
a_1 & 0 & e_{12}^+ & 0 & e_{13}^+ & 0 \\ 0 & a_1 & 0 & e_{12}^- & 0 & e_{13}^- \\
e_{12}^+ & 0 & a_2 & 0 & e_{23}^+ & 0 \\ 0 & e_{12}^- & 0 & a_2 & 0 & e_{23}^- \\
e_{13}^+ & 0 & e_{23}^+ & 0 & a_3 & 0 \\ 0 & e_{13}^- & 0 & e_{23}^- & 0 & a_3
\end{array} \right).
\end{equation}
Each off-diagonal element $e_{ij}^{\pm}$ is shown in Appendix \ref{sec:cmstandardel}. The coefficient $a_i$'s must satisfy the following triangular inequality due to the uncertainty principle:
\begin{equation} \label{eq:triangularcoef}
\left| c_2-c_3 \right| \leq 1 \leq c_2+c_3 , \textrm{ where } c_j = \frac{a_j-\frac{1}{2}}{a_1-\frac{1}{2}} \textrm{ for } j=2,3.
\end{equation}
%Note that two-mode reduced CM of Eq. (\ref{eq:cmstandard3}) becomes directly the standard form (\ref{eq:cmstandard2}).

\section{\label{sec:communication}CV dense coding and single-mode communication}

% Braunstein scheme
\subsection{CV dense coding}
In the CV dense-coding scheme originally proposed by Braunstein and Kimble \cite{bib:PhysRevA.61.042302}, Alice and Bob make use of an initially shared TMSV with squeezing parameter $s$, of which Wigner function is
\begin{equation}
W_{\rho_{AB}} = \frac{1}{\pi^2} \exp\left[ -e^{2s} \left( x_-^2 + p_+^2 \right) -e^{-2s} \left( x_+^2 + p_-^2 \right) \right] ,
\end{equation}
where $ x_\pm = (x_1 \pm x_2)/\sqrt{2} $ and $ p_\pm = (p_1 \pm p_2)/\sqrt{2} $. Alice encodes two classical variables \{$\alpha_x$, $\alpha_p$\} by displacing her mode in phase space with the amplitude $\alpha = \alpha_x + i\alpha_p $. The probability distribution of the encoded amplitude can be taken as 
\begin{eqnarray}\label{eq:dist}
P(\alpha) = \frac{1}{\pi\sigma^2} \exp(-\frac{|\alpha|^2}{\sigma^2}). 
\end{eqnarray}
Then, Alice sends her mode to Bob, who combines it with his mode at a 50/50 beam splitter. Finally, Bob carries out two quadrature measurements, $x$ and $p$, on each of the output modes, respectively. The measurement outcomes are related to the quadrature amplitudes before the beam splitter as $x_-$ and $p_+$. Denoting Bob's measured outcomes as $\{\beta_x,\beta_p\}$, 
the probability distribution $P(\beta|\alpha)$ conditioned on the input $\alpha$ is given by 
\begin{eqnarray}\label{eq:conditional probability}
P(\beta|\alpha)=\int dy_1dx_2 W_{\rho_{AB}'}\left(\beta_x,y_1;x_2,\beta_p\right),
\end{eqnarray}
where $W_{\rho_{AB}'}\left(x_1,y_1;x_2,y_2\right)$ is the Wigner function of the output state at Bob's station, $\rho_{AB}'=U_{\rm BS}D_1(\alpha)\rho_{AB}D_1^\dag(\alpha)U^\dag_{\rm BS}$ with the beam-splitting $U_{\rm BS}$ and the displacement $D_1(\alpha)$ acting on the input state $\rho_{AB}$.
 The achievable information in this scheme can be quantified by the mutual information between the two sets of variables, \{$\alpha_x$, $\alpha_p$\} and $\{\beta_x,\beta_p\}$, 
\begin{eqnarray}
H(A:B)=H(\beta)-H(\beta|\alpha).
\end{eqnarray} 
Here, $H$ denotes the Shannon entropy as
\begin{eqnarray}
H(\beta)&=&-\int d^2\beta P(\beta)\log P(\beta),\nonumber\\
H(\beta|\alpha)&=&-\int d^2\alpha P(\alpha) \int d^2\beta P(\beta|\alpha)\log P(\beta|\alpha),\nonumber\\
\end{eqnarray}
which can be obtained using Eqs. (\ref{eq:dist}) and (\ref{eq:conditional probability}) with $P(\beta)=\int d^2\alpha P(\alpha)P(\beta|\alpha)$. 
 
For a TMSV input, the mutual information turns out to be
\begin{equation}\label{eq:TMSV mutual}
H(A:B) = \ln(1+\sigma^2e^{2s}) .
\end{equation}
It increases with the squeezing parameter $s$, for Bob can read information more accurately due to the decreased variances of the correlated quadrature  $ \langle ( \Delta x_- )^2 \rangle = \langle ( \Delta p_+ )^2 \rangle = \exp(-2s)/2 $. Another point to mention is that the mutual information becomes infinite with $\sigma\rightarrow\infty$, where $\sigma$ is the variance of encoded variable $\alpha$ in Eq.  (\ref{eq:dist}). It is true regardless of squeezing $s$, which is simply the result of communicating infinitely large amount of information. However, it would  require an infinite energy for encoding, thus, the constraint of finite-energy $\bar{n}$ is typically introduced for a fair comparison between different schemes.

When the average number of photons passing through the channel (Alice mode's energy plus the noise added by displacement) is restricted to $\bar n$, one may adjust $\sigma$ in order to maximize the mutual information in Eq.~(\ref{eq:TMSV mutual}) for a given $s$. With $ \bar{n} = \sigma^2 + \sinh^2 s$ for a TMSV, the optimal value occurs at the choice of $ \sigma^2 = \cosh s \sinh s $ and the CV dense coding capacity, i.e. optimized mutual information, is given by
\begin{equation} \label{eq:capacityTMSV}
C_\textrm{dense} = \ln(1+\bar{n}+\bar{n}^2) .
\end{equation}

\subsection{single-mode schemes}
On the other hand, there are three well-known single-mode schemes under the energy constraint $\bar n$. The first one makes use of a coherent state, which is displaced by $\alpha$ (encoding) and then measured via heterodyne detection, i.e. simultaneous measurements of two quadratures (decoding). This coherent-state scheme yields the capacity
\cite{bib:RevModPhys.58.1001}
\begin{equation}\label{eq:capacity coh}
C_\textrm{coh} = \ln(1+\bar{n}) .
\end{equation}
We note that the CV dense-coding scheme using a TMSV always beats the coherent-state scheme regardless of $\bar n>0$ [Cf. Eqs. ~(\ref{eq:capacityTMSV}) and ~(\ref{eq:capacity coh})].

The second scheme employs a squeezed state, which is displaced by the amount $x$ along the squeezed axis (encoding) and measured via homodyne detection (decoding). This Gaussian scheme, which can be a best strategy under the restriction of Gaussian states and Gaussian operations  \cite{bib:RevModPhys.58.1001}, yields the capacity
\begin{equation} \label{eq:capacitysq}
C_\textrm{sq} = \ln(1+2\bar{n}).
\end{equation}
Compared with Eq. ~(\ref{eq:capacityTMSV}), this bound for single-mode Gaussian communication can be beaten by the CV dense coding for $ \bar{n} > 1 $ and it was demonstrated experimentally in Ref. \cite{bib:PhysRevA.71.012304}.

The ultimate capacity of single-mode communication, when non-Gaussian operations are also available, is given by the maximum possible entropy of single-mode state under $\bar{n}$ constraint. This bound, known as the Holevo bound, is achieved when Alice encodes information in number states according to a thermal distribution and Bob decodes the information by photon number counting \cite{bib:PhysRevLett.70.363}, although this scheme requires highly demanding experimental tasks. In this case, the capacity is given by
\begin{equation} \label{eq:capacityFock}
C_\textrm{Fock} = (1+\bar{n})\ln(1+\bar{n}) - \bar{n}\ln\bar{n}.
\end{equation}
The CV dense coding can also beat this bound with a sufficiently large photon number, i.e., $ \bar{n} > 1.8835 $.

\section{\label{sec:twomode}CV dense coding with arbitrary two-mode Gaussian states}

\subsection{Mutual information}
Now we study the CV dense coding protocol using an arbitrary two-mode Gaussian state as a quantum channel beyond TMSVs. 
In the latter case, the conditions to beat the single-mode schemes are given only in terms of energy $\bar n$, just because there exists a monotonic relation between the degree of squeezing (entanglement) and $\bar n$, which is generally not the case. We here intend to derive general conditions to beat the squeezed-state scheme and the photon-number scheme, respectively, for an arbitrary two-mode Gaussian state. 

We first assume that Alice and Bob share a two-mode state having correlations between $x_1$ and $x_2$ and between $p_1$ and $p_2$, respectively, with the condition $ \langle \Delta x_- \Delta p_+ \rangle = 0 $. %without $x-p$ correlation. In other words, if we measure $x_-$ and $p_+$ simultaneously, we have $ \langle \Delta x_- \Delta p_+ \rangle = 0 $. 
In the next section, we show that the scheme becomes optimal under the condition $ \langle \Delta x_- \Delta p_+ \rangle = 0 $, which can always be met via certain local phase-rotations for a given state.

For the case of TMSV, Alice encodes the same amount of information on both quadratures, which is reasonable because two correlated quadratures $x_-$ and $p_+$ have the same variances. In general, however, the communication capacity can be enhanced by encoding a different amount of information on each quadrature as
\begin{equation} \label{eq:encoding}
P(\alpha) = \frac{1}{\pi\sigma_x\sigma_p} \exp \left( -\frac{\alpha_x^2}{\sigma_x^2}-\frac{\alpha_p^2}{\sigma_p^2} \right),
\end{equation}
as will be shown below.
After Alice sends her mode to Bob, Bob measures two correlated quadratures $x_-$ and $p_+$. The conditional probability distribution of Bob reading $\beta$ when Alice encodes $\alpha$ is necessarily a Gaussian distribution that has variances $ V_{x_-} \equiv \langle ( \Delta x_- )^2 \rangle $ and $ V_{p_+} \equiv \langle ( \Delta p_+ )^2 \rangle $ centered at $\alpha/\sqrt{2}$, that is,
\begin{eqnarray} \label{eq:decoding}
P(\beta | \alpha) & = & \frac{1}{\pi\sqrt{V_{x_-}V_{p_+}}} \times \nonumber \\
& & \exp \left[ - \frac{\left( \beta_x - \frac{\alpha_x}{\sqrt{2}} \right)}{V_{x_-}} - \frac{\left( \beta_p - \frac{\alpha_p}{\sqrt{2}} \right)}{V_{p_+}} \right].
\end{eqnarray}
The factor $\sqrt{2}$ dividing $\alpha$ emerges from the output signal reduced by the beam splitter interaction. Using Eqs. (\ref{eq:encoding}) and (\ref{eq:decoding}), we obtain the mutual information as
\begin{equation}
H(A:B) = \frac{1}{2} \ln \left[ \left( 1+\frac{\sigma_x^2}{2V_{x_-}} \right) \left( 1+\frac{\sigma_p^2}{2V_{p_+}} \right) \right].
\end{equation}

Under the energy constraint $ \bar{n} = n_0 + n_s $, where $n_0$ is the mean photon number of Alice's mode before encoding and $ n_s = (\sigma_x^2 + \sigma_p^2)/2 $ is the average number of photons added through displacement, we may adjust the amount of information encoded on each quadrature. If we assume $n_s$ to be large enough, $ n_s > | V_{x_-} - V_{p_+} | $, we find that the optimal encoding is achieved for the choices  $ \sigma_x^2 = n_s+(V_{p_+}-V_{x_-}) $ and $ \sigma_p^2 = n_s+(V_{x_-}-V_{p_+}) $. 
It implies, e.g., that we need to encode more information on the $p$-quadrature ($\sigma_p>\sigma_x$) if the state is more strongly correlated with respect to $p$-quadratures 
($V_{x_-} > V_{p_+}$). 
The optimized mutual information can be written as
\begin{equation} \label{eq:capacity2}
H(A:B) \leq H_\textrm{max} = \ln \frac{\bar{n} - n_0 + \left( V_{x_-}+V_{p_+} \right)}{2\sqrt{V_{x_-}V_{p_+}}}.
\end{equation}
On the other hand, if $n_s$ is small, $ n_s \leq |V_{x_-}-V_{p_+}| $, the single-quadrature encoding on the quadrature possessing a smaller variance is optimal, which is not the case of our interest.

\subsection{Criteria to beat single-mode communications}

Now we examine when our CV dense coding protocol can beat the single-mode Gaussian communication by solving
\begin{eqnarray}\label{eq:f}
f & \equiv & \exp\left(H_\textrm{max}\right) - \exp\left(C_\textrm{sq}\right) \nonumber \\
& = &  \frac{\bar{n} - n_0 + \left( V_{x_-}+V_{p_+} \right)}{2u} - (1+2\bar{n}) > 0 , \nonumber \\
& & \quad \textrm{where } u \equiv \sqrt{V_{x_-}V_{p_+}} .
\end{eqnarray}
In Appendix \ref{sec:criterionproof}, we prove that, if $ u < 1/4 $, one can always find a range of $\bar{n}$ satisfying $ f > 0 $, otherwise $f$ can never be positive.  Therefore, a two-mode Gaussian state can be useful beyond the squeezed-state scheme under the condition
\begin{equation}\label{eq:ourcriterion}
V_{x_-}V_{p_+} < \left(\frac{1}{4}\right)^2.
\end{equation}
%\even though high photon number $\bar{n}$ may be required. 
Note that this criterion is more stringent than the entanglement detection criterion in which the product of correlated variances is bounded by $1/4$ \cite{bib:PhysRevA.60.2752}.

On the other hand, for the CV dense coding scheme to be truly dense coding, it must beat the ultimate single-mode communication with the capacity $C_\textrm{Fock}$ involving non-Gaussian operations. In a large $\bar{n}$ regime, $C_\textrm{Fock}$ in Eq. (\ref{eq:capacityFock}) asymptotically behaves as $ C_\textrm{Fock} \sim \ln(e\bar{n}) $. Thus, looking into
\begin{eqnarray}
&&\exp\left(H_\textrm{max}\right) - \exp\left(C_\textrm{Fock}\right) \nonumber \\
& \approx &  \left(\frac{1}{2u}-e\right)\bar{n}  + \frac{\left( V_{x_-}+V_{p_+} \right)- n_0}{2u} > 0
\end{eqnarray}
we find that the CV dense coding protocol beats any single-mode communications under the condition 
\begin{equation}
V_{x_-}V_{p_+} < \left(\frac{1}{2e}\right)^2
\end{equation}
in a large $\bar n$ regime, which is stricter than the condition in Eq. ~(\ref{eq:ourcriterion}).

%On the other hand, if one finds $ u \geq 1/4 $, it does not mean that the state can never beat single-mode Gaussian communication. Because $V_{x_-}V_{p_+}$ is not invariant under local operations, the state can be transformed to be more useful in CV dense coding by proper local operations, which will be discussed shortly.

\subsection{Improving the mutual information via local unitary operations}

We have previously identified the conditions to beat the single-mode schemes in terms of the product $u \equiv \sqrt{V_{x_-}V_{p_+}}$. This product $u$ is, however, not invariant under local sympletic operations, which makes it possible that the communication capacity can be further improved by local operations for a given state. 
We here investigate how local operations affect the performance of CV dense coding scheme and how one can enhance the mutual information, applying local Gaussian unitary operations, i.e., displacement, phase rotation, and local squeezing. 

\subsubsection{Displacement}
Displacement operation does not affect the second moments but only change the first moments $\bar{\xi}$. The only parameter changed by displacement in Eq. (\ref{eq:capacity2}) is $ n_0 = \langle x_1 \rangle^2 + \langle p_1 \rangle^2 + \langle ( \Delta x_1 )^2 \rangle + \langle ( \Delta p_1 )^2 \rangle $ of Alice mode. To maximize the mutual information, Alice should thus adjust the first moments to zero, $ \langle x_1 \rangle = \langle p_1 \rangle = 0 $ by a proper displacement, to have a minimum $n_0$.

\subsubsection{Phase rotation}
We have derived the mutual information in Eq. ~(\ref{eq:capacity2}) by assuming the condition $ \langle \Delta x_- \Delta p_+ \rangle = 0 $, which can now be relaxed. 
For an arbitrary two-mode state, using Eqs. (\ref{eq:conditional probability}) and (\ref{eq:encoding}), we obtain the mutual information achieved by the CV dense coding as
\begin{widetext}
\begin{equation}
H(A:B) = \frac{1}{2} \ln \left[ 1 + \frac{ \left\langle(\Delta p_+)^2\right\rangle \sigma_x^2 + \left\langle(\Delta x_-)^2\right\rangle \sigma_p^2 + \frac{1}{2}\sigma_x^2\sigma_p^2 }{ 2 \left( \left\langle(\Delta x_-)^2\right\rangle \left\langle(\Delta p_+)^2\right\rangle - \left\langle \Delta x_- \Delta p_+ \right\rangle^2 \right) } \right] .
\end{equation}
This is maximized with the optimal encoding $ \sigma_x^2 = n_s + \left( \langle(\Delta x_-)^2\rangle - \langle(\Delta p_+)^2\rangle \right) $ and $ \sigma_p^2 = n_s + \left( \langle(\Delta p_+)^2\rangle - \langle(\Delta x_-)^2\rangle \right) $, which gives 

\begin{equation} \label{eq:ccapacity}
H(A:B) \leq H_\textrm{max} = \frac{1}{2} \ln \left[ 1 + \frac{ n_s^2 + 2n_s(V_{x_-}+V_{p_+}) + (V_{x_-}-V_{p_+})^2}{4(V_{x_-}V_{p_+}-V_{xp}^2)} \right],
\end{equation}
\end{widetext}
where $V_{xp}\equiv\left\langle \Delta x_- \Delta p_+ \right\rangle$.
 
Now let Alice and Bob perform a phase rotation on their modes so that
\begin{eqnarray}\label{eq:rotation}
\hat{x}_{i}' & = & \hat{x}_{i}\cos\theta_i + \hat{p}_{i}\sin\theta_i , \nonumber \\
\hat{p}_{i}' & = & \hat{p}_{i}\cos\theta_i - \hat{x}_{i}\sin\theta_i,
\end{eqnarray}
($i=1,2$), with $\theta_1=\theta$ and $\theta_2=-\theta$.
The second moments associated with $ \hat{x}_-'$ and $ \hat{p}_+' $ becomes
\begin{eqnarray} \label{eq:rotationvariance}
\left\langle (\Delta x_-')^2 \right\rangle & = & V_{x_-}\cos^2\theta + V_{p_+}\sin^2\theta+\sin2\theta V_{xp}, \nonumber \\
\left\langle (\Delta p_+')^2 \right\rangle & = & V_{p_+}\cos^2\theta + V_{x_-}\sin^2\theta-\sin2\theta V_{xp} , \nonumber \\
\left\langle \Delta x_-' \Delta p_+' \right\rangle & = & V_{xp}\cos2\theta+\frac{1}{2} \left( V_{p_+} - V_{x_-} \right) \sin 2\theta .
\end{eqnarray}
We readily see that the two quantities $V_{x_-}+V_{p_+}$ and $V_{x_-}V_{p_+}-V_{xp}^2$ are invariant under the rotation in Eq. (\ref{eq:rotation}). Furthermore, the optimized 
$H_\textrm{max}$ over the angle $\theta$ occurs at $\tan2\theta=\frac{V_{x_-}-V_{p_+}}{2V_{xp}}$, which leads to $V'_{xp}=\left\langle \Delta x_-' \Delta p_+' \right\rangle=0$. 
Therefore, by a phase rotation, the correlation $V'_{xp}$ can be made zero and an optimal communication capacity arises with the value
\begin{eqnarray} \label{eq:capacity3}
H_\textrm{max}=\ln \frac{\bar{n} - n_0 + \left( V_{x_-}+V_{p_+} \right)}{2\sqrt{V_{x_-}V_{p_+}-V_{xp}^2}}.
\end{eqnarray}
%In this case, the mutual information achieved by CV dense coding can be written as
%\begin{widetext}
%\begin{equation}
%H'(A:B) = \frac{1}{2} \ln \left[ 1 + \frac{ \left\langle(\Delta p_+')^2\right\rangle \sigma_x^2 + \left\langle(\Delta x_-')^2\right\rangle \sigma_p^2 + \frac{1}{2}\sigma_x^2\sigma_p^2 }{ 2 \left( \left\langle(\Delta x_-')^2\right\rangle \left\langle(\Delta p_+')^2\right\rangle - \left\langle \Delta x_-' \Delta p_+' \right\rangle^2 \right) } \right] .
%\end{equation}
%This is maximized with the optimal encoding $ \sigma_x^2 = n_s + \left( \langle(\Delta x_-')^2\rangle - \langle(\Delta p_+')^2\rangle \right) $ and $ \sigma_p^2 = n_s + \left( \langle(\Delta p_+')^2\rangle - \langle(\Delta x_-')^2\rangle \right) $ and the maximum is written by
%\begin{equation}
%H'(A:B) \leq H_\textrm{max}' = \frac{1}{2} \ln \left[ 1 + \frac{ n_s^2 + 2n_s(V_{x_-}+V_{p_+}) + (V_{x_-}-V_{p_+})^2\cos^2 2\theta }{4V_{x_-}V_{p_+}} \right] .
%\end{equation}
%\end{widetext}
Comparing Eq. (\ref{eq:capacity3}) with Eq. (\ref{eq:capacity2}), we now see that a general condition to beat the squeezed-state scheme is given by 
\begin{eqnarray}\label{eq:ccondition}
V_{x_-}V_{p_+}-V_{xp}^2<\left(\frac{1}{4}\right)^2
\end{eqnarray}
That is, if a given state satisfies Eq. (\ref{eq:ccondition}), one readily sees that $H_\textrm{max}$ of Eq. (\ref{eq:ccapacity}) can be larger than $C_\textrm{sq}$ of Eq. (\ref{eq:capacitysq}) in a large $\bar n$ region. On the other hand, if the given state has the correlation as $V_{x_-}V_{p_+}-V_{xp}^2>\left(1/4\right)^2$, the product $V'_{x_-}V'_{p_+}>\left(1/4\right)^2$ is also obtained in a frame of $V'_{xp}=0$, using the invariance $V_{x_-}V_{p_+}-V_{xp}^2$ under our rotation described above, where an optimized capacity arises. However, we have already shown in Appendix B that for a case of $V'_{xp}=0$, the state with the condition $V'_{x_-}V'_{p_+}>\left(1/4\right)^2$ cannot beat the squeezed-state scheme, nor does the given state before rotation.

\subsubsection{Local squeezing}

\begin{figure}[t]
\centering \includegraphics[clip=true, width=0.8\columnwidth]{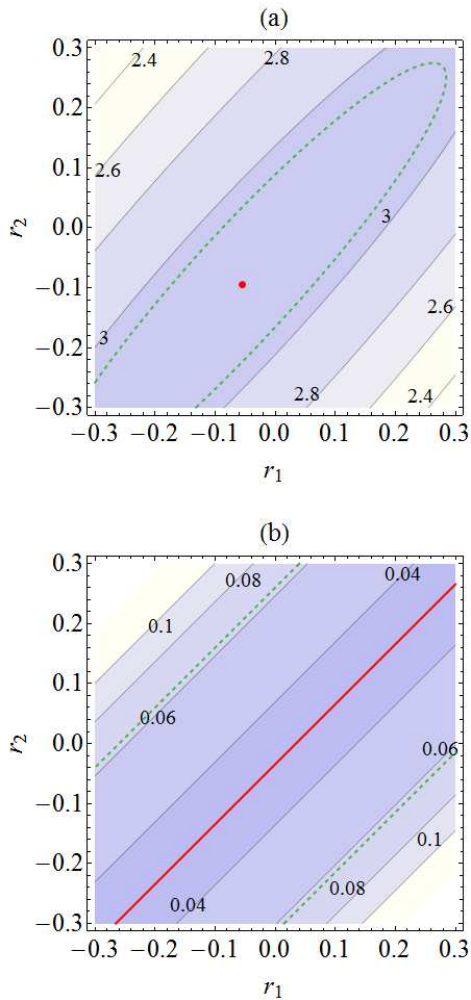}
\caption{\label{fig:squeezing}(a) Plot illustrating the mutual information $H$ against local squeezing parameters $r_1$ and $r_2$. A red dot represents the maximum point and the green dashed curve represents the points of the mutual information equal to the capacity of squeezed-state communication $C_\textrm{sq}$ under the same energy constraint. (b) Plot illustrating $V_{x_-}V_{p_+}$ against squeezing parameters $r_1$ and $r_2$. Red thick line represents the minimum where $ r_1-r_2 = \ln t_\textrm{opt} $ (See main text). The criterion in Eq. \ref{eq:criterion} is satisfied in the region between two green dashed lines. In both (a) and (b), we use the parameters $a_1=1.2, a_2=1.4, a_3=0.9$ and energy constraint $\bar{n}=10$.}
\end{figure}

In the previous subsections, we have shown that an optimal CV dense coding arises under the conditions of zero mean amplitudes and $ \langle \Delta x_- \Delta p_+ \rangle = 0 $. Henceforth, we consider a two-mode CM in standard form (\ref{eq:cmstandard2}) with only local squeezing operations applied to a given state. A symplectic transformation corresponding to a local squeezing is represented by $ S_i = \textrm{diag} (e^{r_i},e^{-r_i}) $ for each mode $i$. The CM after local squeezings is written as
\begin{equation}
\left( \begin{array}{cccc}
a_1 e^{2r_1} & 0 & c_{12} e^{r_1+r_2} & 0 \\ 0 & a_1 e^{-2r_1} & 0 & d_{12} e^{-r_1-r_2} \\ c_{12} e^{r_1+r_2} & 0 & a_2 e^{2r_2} & 0 \\ 0 & d_{12} e^{-r_1-r_2} & 0 & a_2 e^{-2r_2}
\end{array} \right) .
\end{equation}
First let us take an example of two-mode states, which is a two-mode reduced state out of a pure three-mode state in (\ref{eq:cmstandard3}), with $a_1=1.2, a_2=1.4, a_3=0.9$. In Fig. \ref{fig:squeezing}(a), we plot the mutual information against squeezing parameters $r_1$ and $r_2$.
A maximum value (red dot) is found at $ r_1 \approx -0.054 $ and $ r_2 \approx -0.095 $, which means that the standard form before the squeezing operations is not optimal for CV dense coding. In general, it is a nontrivial task to find optimal squeezing parameters $r_1$ and $r_2$ that maximize the mutual information Eq. (\ref{eq:capacity2}), because the quantities $n_0$, $V_{x_-}$, and $V_{p_+}$ are not invariant under local squeezing.

Instead we here optimize $V_{x_-}V_{p_+}$ which is a key quantity in the criterion of Eq. (\ref{eq:criterion}), which can be written as
\begin{equation}
V_{x_-}V_{p_+} = \left( \frac{a_1 t + a_2 t^{-1}}{2} - e_{12}^+ \right) \left( \frac{a_1 t^{-1} + a_2 t}{2} - e_{12}^- \right) ,
\end{equation}
with $ t \equiv \exp(r_1-r_2) $. Note that $V_{x_-}V_{p_+}$ is independent of the sum $r_1+r_2$. By solving $ \frac{d}{dt}(V_{x_-}V_{p_+}) = 0 $, we can find an explicit expression of $t_\textrm{opt}$ that minimizes $V_{x_-}V_{p_+}$, although the expression is lengthy. In Fig. \ref{fig:squeezing}(b), we plot $V_{x_-}V_{p_+}$ against squeezing parameters for the case we considered above. We see that $V_{x_-}V_{p_+}$ has the same value along the line $ r_1-r_2 = \textrm{(const)} $. The minimum is achieved when $ r_1-r_2 \approx -0.034 $, which is not the case of a standard form. 

\begin{figure}
\centering \includegraphics[clip=true, width=0.8\columnwidth]{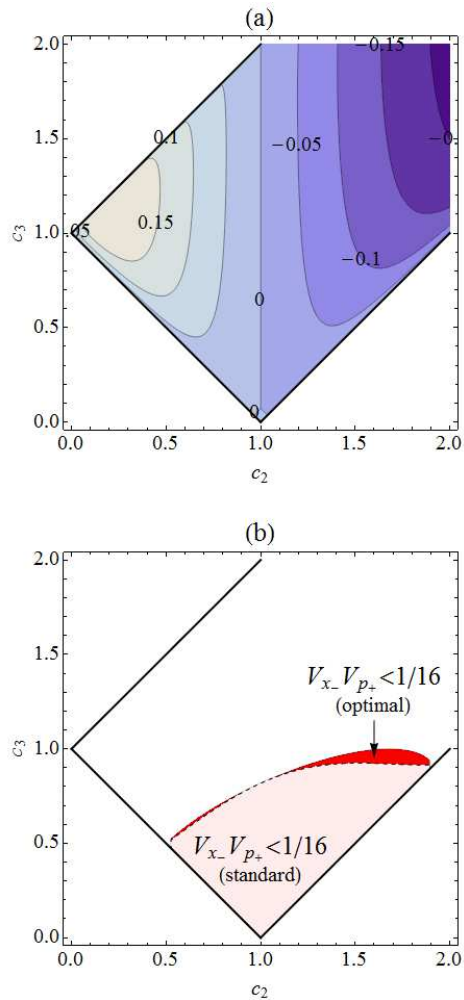}
\caption{\label{fig:optimization}(a) Plot of optimal $r_1-r_2$ (b) Plot illustrating the region where the criterion (\ref{eq:criterion}) is satisfied with(without) local squeezing. We use the parameter $a_1=1.5$ and $a_{2(3)}$ is rescaled in terms of $c_{2(3)}$. Thick lines represent the boundary of pure quantum states, given by Eq. (\ref{eq:triangularcoef}).}
\end{figure}

In Fig. \ref{fig:optimization}(a), we plot the optimal $r_1-r_2$ for the two-mode reduced state of a pure three-mode state in standard form (\ref{eq:cmstandard3}) with a fixed $a_1(=1.5)$ and varying $a_2$ and $a_3$ (scaled in terms of $c_2$ and $c_3$).
For a symmetric state with $ a_1 = a_2 $ ($ c_2 = 1 $), $V_{x_-}V_{p_+}$ is minimized when $ r_1-r_2 = 0 $ so that the standard form is optimal. Otherwise, a nonzero local squeezing is required for optimization. By minimizing $V_{x_-}V_{p_+}$, we can modify a state which does not initially satisfy the criterion of Eq. (\ref{eq:criterion}) to a form satisfying it via local squeezing. In Fig. \ref{fig:optimization}(b), we plot the region where the criterion (\ref{eq:criterion}) is satisfied with(without) local squeezing with a fixed $a_1(=1.5)$ and varying $a_2$ and $a_3$. We find the region between the solid curve and the dashed curve, where $ V_{x_-}V_{p_+} < 1/16 $ is satisfied with optimization but not in the standard form.

\section{\label{sec:multimode}CV dense coding with multi-mode Gaussian states}

% plot without optimization
In this section, we extend our study of CV dense coding to the case of multipartite Gaussian states, specifically employing different pairs of two modes out of multiparties. Let us first assume that a pure three-mode Gaussian state $\rho_0$ is distributed to three parties, Alice, Bob, and Charlie in the standard form of Eq. (\ref{eq:cmstandard3}). Alice here plays as an information sender and either Bob or Charlie is a receiver. 

\begin{figure}[t]
\centering \includegraphics[clip=true, width=0.8\columnwidth]{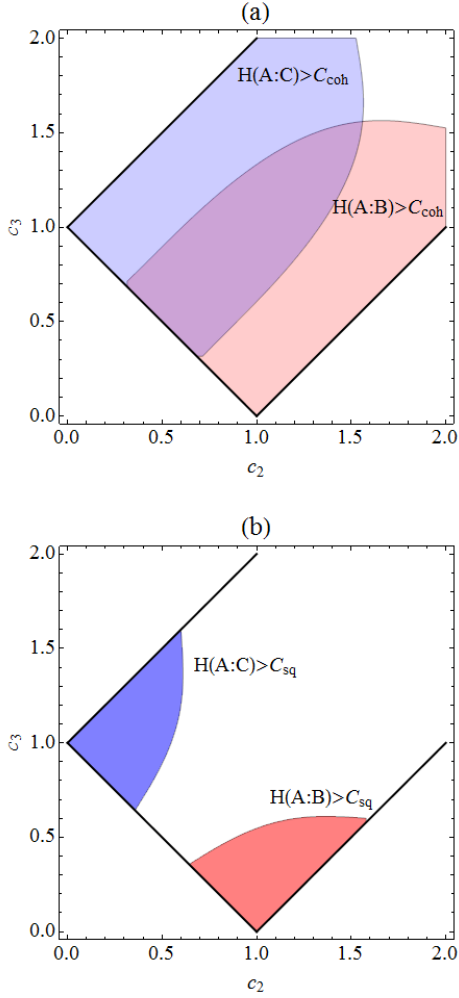}
\caption{\label{fig:capacity}Plot illustrating the regions where the mutual information surpasses the capacity of (a) coherent-state communication and of (b) squeezed-state communication. We use the parameter $ a_1 = 1.5 $, and $a_2$ and $a_3$ are rescaled in terms of $c_2$ and $c_3$. Energy constraint is given by $ \bar{n} = 10 $. Thick lines represent the boundary of pure quantum states, given by Eq. (\ref{eq:triangularcoef}).}
\end{figure}

First we consider the case without optimization and obtain the mutual information $H(A:B)$ and $H(A:C)$, respectively, for a given three-mode state. In Fig. \ref{fig:capacity}, we plot the region where the mutual information attained by CV dense coding protocol between Alice and Bob (Charlie), $H(A:B(C))$, surpasses the capacity of Gaussian single-mode schemes.
We see that the CV dense coding protocol beats the coherent-state communication in a broad region and that both $H(A:B)$ and $H(A:C)$ can surpass the capacity of coherent-state communication simultaneously in a certain region. However, we do not find the region where both $H(A:B)$ and $H(A:C)$ surpass the capacity of squeezed-state communication simultaneously. It means that Alice can not make a useful CV dense coding to beat the squeezed-state communication with both receivers simultaneously.

\begin{figure}[t]
\centering \includegraphics[clip=true, width=0.7\columnwidth]{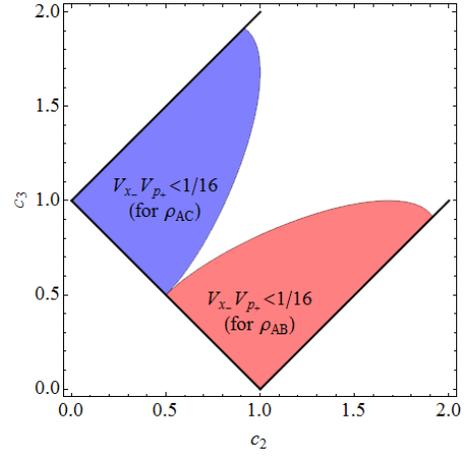}
\caption{\label{fig:criterion}Plot illustrating region where the criterion (\ref{eq:criterion}) is satisfied for $\rho'_{AB}$(red) and $\rho''_{AC}$(blue) with optimal local squeezings. We use the parameter $ a_1 = 1.5 $, and $a_2$ and $a_3$ are rescaled in terms of $c_2$ and $c_3$. Thick lines represent the boundary of pure quantum states, given by Eq. (\ref{eq:triangularcoef}).}
\end{figure}

% monogamy relation with optimization
We now show that even if they attempt to optimize the mutual information by applying local unitary operations for the two pairs \{A,B\} and \{A,C\}, respectively, it is impossible to beat the squeezed-state communication with both receivers simultaneously. Alice and Bob (Charlie) now perform optimal local squeezing operations on their modes and have the optimized state $\rho'_{AB}$ ($\rho''_{AC}$). We prove that the criterion to beat squeezed-state communication, Eq. (\ref{eq:criterion}), cannot be satisfied for both $\rho'_{AB}$  and $\rho''_{AC}$ because the product of variances is bounded, regardless of local squeezing parameters, as
\begin{eqnarray}
& & \left\langle (\Delta x_-)^2 \right\rangle_{\rho'_{AB}} \left\langle (\Delta p_+)^2 \right\rangle_{\rho'_{AB}} \left\langle (\Delta x_-)^2 \right\rangle_{\rho''_{AC}} \left\langle (\Delta p_+)^2 \right\rangle_{\rho''_{AC}} \nonumber \\
& & \geq \left( \frac{1}{16} \right)^2 \left| \left\langle \left[ \hat{x}_1 , \hat{p}_1 \right] \right\rangle_{\rho_0} \right|^4 \geq \left(\frac{1}{16}\right)^2 .
\end{eqnarray}
The main idea of proof is to apply Heisenberg's uncertainty principle for non-commuting operators $\hat{x}_{1}-\hat{x}_{2}$ and $\hat{p}_{1}+\hat{p}_{3}$ and also for $\hat{x}_{1}-\hat{x}_{3}$ and $\hat{p}_{1}+\hat{p}_{2}$ (see Appendix \ref{sec:variancebound} for details). Therefore, we find a strict monogamy relation for CV dense coding: if Alice has quantum advantage in CV dense coding with Bob, she can never have quantum advantage with Charlie. In Fig. \ref{fig:criterion}, we plot the region where $ V_{x_-}V_{p_+} < 1/16 $ is satisfied for $\rho'_{AB}$ and $\rho''_{AC}$.
We see that there is no overlap between two regions so that a strict monogamy relation is satisfied. As the condition to beat the Holevo bound (number-state scheme) is even stricter as $ V_{x_-}V_{p_+} < 1/4e^2 $, we also conclude that a true dense-coding is possible only for a single pair of two modes.

% mixed state
% generalization to multi-mode state
A general mixed three-mode state may not be written in the standard form (\ref{eq:cmstandard3}). However, if the three-mode state has the condition $ \langle \Delta x_- \Delta p_+ \rangle = 0 $ for both pairs of \{A,B\} and \{A,C\}, the proof in Appendix \ref{sec:variancebound} is still valid, which in fact applies to any two pairs of two modes selected out of many parties beyond three-mode cases. 

For a symmetric, mixed, $N$-mode state which is invariant under the permutation of modes, we generally obtain a strict monogamy relation without resort to the condition $ \langle \Delta x_- \Delta p_+ \rangle = 0 $. In Sec. IV C 2, we have derived a general conditon to beat the squeezed-state scheme, $V_{x_-}V_{p_+}-V_{xp}^2<\left(1/4\right)^2$ in Eq. (\ref{eq:ccondition}). As shown before, it is equivalent to $V'_{x_-}V'_{p_+}<\left(1/4\right)^2$ eliminating $V_{xp}$ by a rotation $\{\theta,-\theta\}$ of each mode, respectively. Let $\{\theta,-\theta\}$ and $\{\theta',-\theta'\}$ denote those angles of rotation for each pair of two modes \{A,B\} and \{A,C\}, respectively. For a symmetric state, we have $\theta=\theta'$ and prove that $V'_{x_-}V'_{p_+}<\left(1/4\right)^2$ is possible for only one pair in Appendix \ref{sec:variancebound}.

%However, if Alice needs to do phase rotation operation to eliminate $x-p$ correlation, we cannot directly apply the proof in Appendix \ref{sec:variancebound}. Nonetheless, we conjecture that monogamy relation is satisfied for any mixed three-mode state because, in most cases, pure states carry out quantum protocols more efficiently than mixed states. Monogamy relation for mixed three-mode state can be generalized to that for $N$-mode state. Because one sender can not have quantum advantage with two receivers simultaneously for any reduced three-mode state, she can have quantum advantage with only one receiver among $N-1$ receivers.

\section{\label{section:conclusion}Conclusion}

% summary
In this paper, we have studied the CV dense-coding protocol employing an arbitrary two-mode Gaussian state. We have particularly derived criteria, Eq. (1) or its generalized form  Eq. (\ref{eq:ccondition}), to detect two-mode states that can be more useful for communication than single-mode schemes, namely the squeezed-state scheme (best among Gaussian schemes) and the number-state scheme (the optimal scheme achieving the Holevo bound). We have also shown how to enhance the performance of CV dense coding for a given two-mode state via local operations. 

We have also extended our study to a multipartite Gaussian state and proved that there exists a strict monogamy relation among different parties in view of the operational entanglement, i.e., communication capacity via dense-coding protocol. That is, a sender (Alice) can have quantum advantage over the single-mode schemes strictly with one receiver only, which was proved for the case of symmetric $N$-mode systems unconditionally and for the case of general multimode systems possessing $\{x_i,x_j\}$- and $\{p_i,p_j\}$-correlations only.

% teleportation fidelity
For a future work, it will be interesting to study the CV dense-coding and its monogamy relation beyond Gaussian regime. It has been known that some non-Gaussian operations can enhance performance of quantum information tasks, e.g. teleportation \cite{PhysRevA.61.032302}. Thus, it is of fundamental and practical interest whether the non-Gaussian regime can manifest a different feature from the monogamy relation identified here. In addition, the study of monogamy relation can be further extended to other operational measures like the output fidelity of CV teleportation. These studies may contribute to our understanding of multipartite entanglement structure in CV systems providing an insight into quantum network communication using multimode CV systems.

\section*{acknowledgement}
This work is supported by the NPRP grant 4-554-1-084 from Qatar National Research Fund.

\onecolumngrid
\appendix

\section{Standard form CM of pure three-mode Gaussian states} \label{sec:cmstandardel}

A pure three-mode Gaussian state satisfies
\begin{equation}
\textrm{Det}\boldsymbol{\sigma}_{ABC} = \left( \frac{1}{2} \right)^3, ~ \textrm{Det}\boldsymbol{\sigma}_{ij} = \left( \frac{1}{2} \right)^2 \textrm{Det}\boldsymbol{\sigma}_{k} ,
\end{equation}
where $\boldsymbol{\sigma}_{ij}$ ($\boldsymbol{\sigma}_{k}$) is the reduced two-mode (single-mode) CM of $\boldsymbol{\sigma}_{ABC}$ with $i,j,k=1,2,3$ different from each other. From these conditions, after some algebra, one finds the expression of off-digonal elements of CM in Eq. (\ref{eq:cmstandard3}) as
\begin{equation}
e_{ij}^{\pm} \equiv \frac{\sqrt{[(a_i-a_j)^2-(a_k-\frac{1}{2})^2][(a_i-a_j)^2-(a_k+\frac{1}{2})^2]}\pm\sqrt{[(a_i+a_j)^2-(a_k-\frac{1}{2})^2][(a_i+a_j)^2-(a_k+\frac{1}{2})^2]}}{4\sqrt{a_i a_j}} .
\end{equation}

\section{Proof for the criterion of Equation (\ref{eq:criterion})} \label{sec:criterionproof}

Arranging $f$ in Eq. (\ref{eq:f}), we have
\begin{equation} \label{eq:ff}
f = \frac{1-4u}{2u}n_s + \frac{V_{x_-}+V_{p_+}}{2u} - (1+2n_0) .
\end{equation}

\begin{itemize}

\item {\bf Case I:} $ u < 1/4 $ 

For a given state, the last two terms in the right-hand side of Eq. (\ref{eq:ff}) are finite. Thus, we can always find $f>0$ with a sufficiently large $n_s$.

\item {\bf Case II:} $ u \geq 1/4 $ 

Above all, we derive the bound for $V_{x_-}$. Without loss of generality, let us assume $ V_{x_-} \leq V_{p_+} $. A lower bound arises from Heisenberg's uncertainty principle as
\begin{eqnarray}
& & \left\langle \Delta^2 \hat{x}_- \right\rangle \left\langle \Delta^2 \hat{p}_1 \right\rangle \geq \frac{1}{4} \left| \left\langle \left[ \frac{\hat{x}_1-\hat{x}_2}{\sqrt{2}} , \hat{p}_1 \right] \right\rangle \right|^2 = \frac{1}{8} \left| \left\langle \left[ \hat{x}_1 , \hat{p}_1 \right] \right\rangle \right|^2 = \frac{1}{8} , \nonumber \\
& & \qquad \textrm{or } ~ V_{x_-} \geq \frac{1}{8\left\langle \Delta^2 \hat{p}_1 \right\rangle} .
\end{eqnarray}
Taking partial derivative of $H_\textrm{max}$ with respect to $V_{p_+}$, we find
\begin{equation}
\frac{\partial H_\textrm{max}}{\partial V_{p_+}} = -\frac{n_s+V_{x_-}-V_{p_+}}{V_{p_+}(n_s+V_{x_-}+V_{p_+})} ,
\end{equation}
which is negative under the two-quadrature encoding condition $ n_s > \left|V_{x_-}-V_{p_+}\right| $. Therefore $H_\textrm{max}$ becomes maximum when $V_{p_+}$ is minimum, i.e. $ V_{p_+} = V_{x_-} $, which is given by
\begin{equation}
H_\textrm{max} \leq \ln\frac{n_s+2V_{x_-}}{2V_{x_-}} .
\end{equation}
In order to beat squeezed-state communication, $V_{x_-}$ must be smaller than $\frac{n_s}{4\bar{n}}$.

On the other hand, we can find the upper bound of $f$ under the encoding condition $ n_s > \left|V_{x_-}-V_{p_+}\right| $:
\begin{eqnarray}  \label{eq:fbound}
f & < & \frac{1-4u}{2u}\left(V_{p_+}-V_{x_-}\right) + \frac{V_{x_-}+V_{p_+}}{2u} - (1+2n_0) \nonumber \\
& = & -2\left(\sqrt{V_{p_+}}-\frac{1}{4\sqrt{V_{x_-}}}\right)^2 + \frac{1}{8V_{x_-}} + 2V_{x_-} -  (1+2n_0) \nonumber \\
& \leq & g(V_{x_-}) - (1+2n_0) , \qquad \textrm{where } ~ g(x) \equiv \frac{1}{8x}+2x .
\end{eqnarray}
Since $g(x)$ is a decreasing function in the range $ 0 < x < 1/4 $, we find the range of $g(V_{x_-})$ as
\begin{equation}
g\left(\frac{n_s}{4\bar{n}}\right) < g\left(V_{x_-}\right) \leq g\left(\frac{1}{8\left\langle \Delta^2 \hat{p}_1 \right\rangle}\right)
\end{equation}
The upper bound satisfies
\begin{equation}
g\left(\frac{1}{8\left\langle \Delta^2 \hat{p}_1 \right\rangle}\right) = \left\langle \Delta^2 \hat{p}_1 \right\rangle + \frac{1}{4\left\langle \Delta^2 \hat{p}_1 \right\rangle} \leq \left\langle \Delta^2 \hat{p}_1 \right\rangle + \left\langle \Delta^2 \hat{x}_1 \right\rangle = 1+2n_0 .
\end{equation}
where the Heisenberg-uncertainty inequality is used. By putting the upper bound of $g(V_{x_-})$ into Eq. (\ref{eq:fbound}), we obtain $ f < 0 $.

\end{itemize}

\section{Lower bound for the product of correlated variances} \label{sec:variancebound}

With operators $\hat{x}_i$ and $\hat{p}_i$ for the initial three-mode state $\rho_{0}$, we can write, in the Heisenberg picture, operators for the optimal state $\rho'_{AB}$ as $ \hat{x}_1' = e^{-r_1'}\hat{x}_1 , \hat{p}_1' = e^{r_1'}\hat{p}_1 , \hat{x}_2' = e^{-r_2'}\hat{x}_2 , \hat{p}_2' = e^{r_2'}\hat{p}_1 $, and similarly, $ \hat{x}_1'' = e^{-r_1''}\hat{x}_1 , \hat{p}_1'' = e^{r_1''}\hat{p}_1 , \hat{x}_3'' = e^{-r_3''}\hat{x}_3 , \hat{p}_3'' = e^{r_3''}\hat{p}_3 $ for $\rho''_{AC}$. Then we find 
\begin{eqnarray}
& & \left\langle (\Delta x_-)^2 \right\rangle_{\rho'_{AB}} \left\langle (\Delta p_+)^2 \right\rangle_{\rho'_{AB}} \left\langle (\Delta x_-)^2 \right\rangle_{\rho''_{AC}} \left\langle (\Delta p_+)^2 \right\rangle_{\rho''_{AC}} \nonumber \\
& & = \left\langle \Delta^2 \left( \frac{\hat{x}_1'-\hat{x}_2'}{\sqrt{2}} \right) \right\rangle_{\rho_0} \left\langle \Delta^2 \left( \frac{\hat{p}_1'+\hat{p}_2'}{\sqrt{2}} \right) \right\rangle_{\rho_0} \left\langle \Delta^2 \left( \frac{\hat{x}_1''-\hat{x}_3''}{\sqrt{2}} \right) \right\rangle_{\rho_0} \left\langle \Delta^2 \left( \frac{\hat{p}_1''+\hat{p}_3''}{\sqrt{2}} \right) \right\rangle_{\rho_0} \nonumber \\
& & \geq \left( \frac{1}{4} \right)^2 \left| \left\langle \left[ \frac{\hat{x}_1'-\hat{x}_2'}{\sqrt{2}} , \frac{\hat{p}_1''+\hat{p}_3''}{\sqrt{2}} \right] \right\rangle_{\rho_0} \right|^2 \times \left| \left\langle \left[ \frac{\hat{p}_1'+\hat{p}_2'}{\sqrt{2}} , \frac{\hat{x}_1''-\hat{x}_3''}{\sqrt{2}} \right] \right\rangle_{\rho_0} \right|^2 \nonumber \\
& & = \left( \frac{1}{16} \right)^2 \left| \left\langle \left[ e^{-r_1'}\hat{x}_1 , e^{r_1''}\hat{p}_1 \right] \right\rangle_{\rho_0} \right|^2 \times \left| \left\langle \left[ e^{r_1'}\hat{p}_1 , e^{-r_1''}\hat{x}_1 \right] \right\rangle_{\rho_0} \right|^2 \nonumber \\
& & = \left( \frac{1}{16} \right)^2 \left| \left\langle \left[ \hat{x}_1 , \hat{p}_1 \right] \right\rangle_{\rho_0} \right|^4 \geq \left( \frac{1}{16} \right)^2 .
\end{eqnarray}

In the case of applying local phase rotations, we have  
$\hat{x}_{i}' = \hat{x}_{i}\cos\theta_i + \hat{p}_{i}\sin\theta_i$ and $\hat{p}_{i}' = \hat{p}_{i}\cos\theta_i - \hat{x}_{i}\sin\theta_i$.
 Let us take the angles of rotation $\{\theta_1,-\theta_1\}$ and $\{\theta_2,-\theta_2\}$ for each pair of two modes \{A,B\} and \{A,C\}. Then, the 4th line of the above derivation is changed to
 
\begin{eqnarray}
& & = \left( \frac{1}{16} \right)^2 \left| \left\langle \left[ \hat{x}_{1}\cos\theta_1 + \hat{p}_{1}\sin\theta_1, \hat{p}_{1}\cos\theta_2 - \hat{x}_{1}\sin\theta_2 \right] \right\rangle_{\rho_0} \right|^2 \times \left| \left\langle \left[ \hat{p}_{1}\cos\theta_1 - \hat{x}_{1}\sin\theta_1 , \hat{x}_{1}\cos\theta_2 + \hat{p}_{1}\sin\theta_2 \right] \right\rangle_{\rho_0} \right|^2 \nonumber \\
& & = \left( \frac{1}{16} \right)^2 \cos^4(\theta_1-\theta_2)\left| \left\langle \left[ \hat{x}_1 , \hat{p}_1 \right] \right\rangle_{\rho_0} \right|^4 \geq \left( \frac{1}{16} \right)^2 \cos^4(\theta_1-\theta_2).
\end{eqnarray}   

The bound again becomes $\left( \frac{1}{16} \right)^2$  for $\theta_1=\theta_2$.

\end{document}